\documentclass[12pt]{article}   
\usepackage{epsfig}
\newcommand{\mysection}{\setcounter{equation}{0}\section}

\def\beq{\begin{equation}}
\def\eeq{\end{equation}}
\def\beqa{\begin{eqnarray}}
\def\eeqa{\end{eqnarray}}

\newlength{\dinwidth} \newlength{\dinmargin}
\begin{document}
\begin {flushright}
Cavendish-HEP-03/26\\
\end {flushright} 
\vspace{3mm}
\begin{center}
{\Large \bf $W$ hadroproduction at large transverse momentum 
beyond next-to-leading order}
\end{center}
\vspace{2mm}
\begin{center}
{\large Nikolaos Kidonakis and Agust{\' \i}n Sabio Vera}\\
\vspace{2mm}
{\it Cavendish Laboratory\\ University of Cambridge\\
Madingley Road\\ Cambridge CB3 0HE, UK}
\end{center}

\begin{abstract}
We study the production of $W$ bosons at large transverse momentum
in $p{\bar p}$ collisions. We show that the next-to-leading order cross section
at large transverse momentum is dominated by threshold 
soft-gluon corrections.
We add next-to-next-to-leading-order soft-gluon corrections
to the exact next-to-leading-order differential cross sections. 
We find that these higher-order corrections provide modest enhancements 
to the transverse momentum distribution of the $W$ at the Tevatron,
and reduce significantly the dependence on the factorization 
and renormalization scales.
 
\end{abstract}

\thispagestyle{empty} \newpage \setcounter{page}{2}

\mysection{Introduction}

The production of $W$ bosons  in hadron colliders  
is a process of relevance in testing the Standard Model and 
in estimates of backgrounds to new physics.
For example,
$Wb{\bar b}$ production is the principal background to the associated 
Higgs boson production $p{\bar p}\rightarrow H(\rightarrow b{\bar b})W$ 
at the Tevatron \cite{EV}. 

The calculation of the complete next-to-leading-order (NLO)
cross section for $W$ hadroproduction at large transverse momentum was 
presented in Refs. \cite{AR,gpw}.
The NLO results in \cite{AR,gpw} showed an enhancement of the 
differential distributions in transverse momentum $Q_T$ of the $W$ boson.
The $Q_T$ distribution falls rapidly with increasing $Q_T$, spanning five
orders of magnitude in the 30 GeV $< Q_T <$ 190 GeV region 
at the Tevatron. The NLO corrections also considerably stabilize
the dependence of the cross section on the factorization and renormalization
scales.

The calculation of hard-scattering cross sections near partonic threshold, 
such as electroweak-boson
production at high transverse momentum, involves corrections from the 
emission of soft gluons from the partons in the process. 
At each order in perturbation theory one encounters large 
logarithms that arise from incomplete cancellations near partonic
threshold between graphs with real emission and virtual graphs.
This is due to the limited phase space available for real gluon emission
near partonic threshold. These threshold corrections, calculated
in the eikonal approximation, exponentiate as a result of the factorization 
properties \cite{KS,KOS,LOS} of the cross section. 
The cross section is factorized
into functions that describe gluons collinear to the incoming partons,
hard quanta, and noncollinear soft-gluons. The renormalization
group properties of these functions result in resummation.
Such threshold corrections have by now been successfully resummed for many
processes \cite{NK}.

In this paper we further the study of the resummation of threshold logarithms
and the expansion of the resummed cross section at
next-to-next-to-leading order (NNLO)
for electroweak boson hadroproduction, which was first
presented in Ref. \cite{NKVD} at next-to-next-to-leading logarithmic (NNLL) 
accuracy.  
In Ref. \cite{NKVD} no numerical phenomenological studies were made. 
In this paper we study the significance of NNLO soft-gluon corrections
for $W$-boson production at large transverse momentum.
Furthermore we increase the accuracy of the theoretical calculation to
next-to-next-to-next-to-leading logarithms (NNNLL). 
Related studies for direct photon production, including numerical
results, were made in Refs. \cite{NKJO1,NKJO2}.
The partonic subprocesses involved in direct photon production
are similar to the ones discussed in this paper.

Recently, a unified approach to the calculation of NNLO soft and
virtual corrections for processes in hadron-hadron
and lepton-hadron colliders has been presented in Ref. \cite{NKuni}.
This work unifies and extends previous approaches by going beyond
NNLL accuracy.
We follow this reference and calculate the NNLO soft corrections
to $W$ production at large transverse momentum at the Tevatron.
The form of our theoretical results differs slightly from 
(but is consistent with) the one in Ref. \cite{NKVD}, partly in that here 
we use the transverse momentum $Q_T$ as the hard scale instead 
of the mass of the $W$.

We note that there has also been work on resummation of Sudakov logarithms
at small transverse momentum for electroweak boson production
\cite{ERV,KuSt} as well as on joint resummation of $Q_T$ and threshold
logarithms \cite{KSV}.

In Section 2 we discuss some of the kinematics of the partonic
subprocesses involved and we define the corrections to be calculated.
In Section 3 we provide  results for the NLO soft and virtual, and 
the  NNLO soft-gluon corrections.
The NLO soft corrections agree near partonic threshold with the 
exact NLO calculations in Refs. \cite{AR,gpw} while our 
NNLO soft-gluon corrections provide new predictions.
In Section 4 we study the numerical effect of the NNLO soft-gluon
corrections to $W$ hadroproduction at the Tevatron Run I and II,
while also showing that the NLO cross section is dominated by soft-gluon 
corrections.

\mysection{Kinematics and plus distributions}

For the hadronic production of a $W$ boson, with mass $m_W$,
\beq
h_A(P_A)+h_B(P_B) \longrightarrow W(Q) + X \, ,
\eeq
we can write the factorized single-particle-inclusive cross section as
\beqa
E_Q\,\frac{d\sigma_{h_Ah_B{\rightarrow}W(Q)+X}}{d^3Q} &=&
\sum_{f} \int dx_1 \, dx_2 \; \phi_{f_a/h_A}(x_1,\mu_F^2) 
\; \phi_{f_b/h_B}(x_2,\mu_F^2) 
\nonumber \\ && \hspace{-10mm} \times \,
E_Q\,\frac{d\hat{\sigma}_{f_af_b{\rightarrow}W(Q)+X}}{d^3Q}
(s,t,u,Q,\mu_F,\alpha_s(\mu_R^2)) \label{factor}
\label{factW}
\eeqa
where $E_Q=Q^0$, $\phi_{f/h}$ is the parton distribution for parton 
$f$ in hadron $h$,
and $\hat{\sigma}$ is the perturbative parton-level cross section. 
The initial-state collinear singularities are factorized into the
parton distributions at factorization scale $\mu_F$, while $\mu_R$ is the
renormalization scale.

At the parton level, the lowest-order subprocesses for the 
production of a $W$ boson and a parton are 
\beqa
q(p_a)+g(p_b) &\longrightarrow& W(Q) + q(p_c)  \, ,
\nonumber \\
q(p_a)+{\bar q}(p_b) &\longrightarrow& W(Q) + g(p_c)  \, .
\label{partsub}
\eeqa
The hadronic and partonic kinematical invariants in the process are 
\beqa
&& \hspace{-5mm} S=(P_A+P_B)^2, \; T=(P_A-Q)^2, \; U=(P_B-Q)^2, \;
S_2 = S + T + U - Q^2,
\nonumber \\ 
&& \hspace{-5mm} s=(p_a+p_b)^2, \; t=(p_a-Q)^2, \; u=(p_b-Q)^2, \;
s_2 = s + t + u - Q^2,
\label{partkin} 
\eeqa
where $S_2$ and $s_2$ are the invariant masses of the system recoiling 
against the electroweak boson at the hadron and parton levels, respectively.
$s_2=(p_a+p_b-Q)^2$ parametrizes the inelasticity of the parton scattering, 
taking the value $s_2=0$ for one-parton production.
Since $x_i$ is the initial-parton momentum fraction, defined
by $p_a = x_1 P_A$ and $p_b = x_2 P_B$, the hadronic and partonic kinematical 
invariants are related by 
$s = x_1 x_2 S$, $t-Q^2 = x_1 (T-Q^2)$, and $u-Q^2 = x_2 (U-Q^2)$.

In general, the partonic cross section $\hat{\sigma}$ 
includes distributions with respect 
to $s_2$ at $n$-th order in the strong coupling $\alpha_s$ of the type
\beq
\left[\frac{\ln^{m}(s_2/Q_T^2)}{s_2} \right]_+, \hspace{10mm} m\le 2n-1\, ,
\label{zplus}
\eeq
defined by their integral with any smooth function $f$ by 
\beqa
\int_0^{s_{2 \, max}} ds_2 \, f(s_2) \left[\frac{\ln^m(s_2/Q_T^2)}
{s_2}\right]_{+} &\equiv&
\int_0^{s_{2\, max}} ds_2 \frac{\ln^m(s_2/Q_T^2)}{s_2} [f(s_2) - f(0)]
\nonumber \\ &&
{}+\frac{1}{m+1} \ln^{m+1}\left(\frac{s_{2\, max}}{Q_T^2}\right) f(0) \, .
\label{splus}
\eeqa
These ``plus'' distributions are the remnants of cancellations between
real and virtual contributions to the cross section. 
Note that in Ref. \cite{NKVD} we used $Q$ instead of $Q_T$ in the
plus distributions. Here we prefer to use $Q_T$ as we find it a slightly
better physical hard scale for the $Q_T$ distributions that we will
be calculating.    
Below we will make use of the terminology that at $n$-th order in $\alpha_s$
the leading logarithms (LL) 
are those with $m=2n-1$ in Eq. (\ref{zplus}), next-to-leading logarithms (NLL)
with $m=2n-2$, next-to-next-to-leading logarithms (NNLL) with $m=2n-3$,
and next-to-next-to-next-to-leading logarithms (NNNLL) with $m=2n-4$.

\mysection{Next-to-next-to-leading order soft-gluon corrections}

In this section, for each of the subprocesses in Eq. (\ref{partsub})
we first present the NLO soft and virtual corrections.
We then present the NNLO soft-gluon corrections. 
We work in the ${\overline{\rm MS}}$ scheme throughout.

\subsection{The $qg \longrightarrow Wq$ subprocess}

The Born differential cross section for this process is
\beq
E_Q \frac{d\sigma^B_{qg\rightarrow Wq}}{d^3Q}
=F^B_{qg \rightarrow Wq} \, \delta(s_2) \, ,
\eeq
where
\beqa
F^B_{qg \rightarrow Wq} &=& \frac{\alpha \, 
\alpha_s(\mu_R^2)C_F}{s(N_c^2-1)}
A^{qg} \, \sum_{f} |L_{ff_a}|^2 \, ,\\
A^{qg} &=& - \left(\frac{s}{t}+\frac{t}{s}+\frac{2uQ^2}{st}\right) \, , 
\nonumber
\eeqa
with $L$ the left-handed couplings of the
$W$ boson to the quark line, $f$ the quark flavor and $\sum_f$
the sum over the flavors allowed by the CKM mixing and by the energy
threshold. For the $L$ couplings we choose the conventions of
Ref.~\cite{gpw}. Also $C_F=(N_c^2-1)/(2N_c)$ with $N_c=3$ the number
of colors.

We can write the NLO soft and virtual corrections for 
$qg \longrightarrow Wq$ 
in single-particle inclusive kinematics as
\beq
E_Q\frac{d{\hat\sigma}^{(1)}_{qg \rightarrow Wq}}{d^3Q} = 
F^B_{qg \rightarrow Wq} 
{\alpha_s(\mu_R^2)\over\pi}\,
\left\{c_3^{qg} \, \left[\frac{\ln(s_2/Q_T^2)}{s_2}\right]_+ 
+c_2^{qg} \, \left[\frac{1}{s_2}\right]_+ + c_1^{qg} \, 
\delta(s_2)\right\} \, .
\label{qgnlo}
\eeq

Note that the $[\ln(s_2/Q_T^2)/s_2]_+$ term (which is the LL, since
$m=n=1$ in Eq.~(\ref{zplus}))
and the $[1/s_2]_+$ term (NLL, since $n=1$, $m=0$) 
are the soft gluon corrections.
The $\delta(s_2)$ term is the contribution from the virtual corrections.
Below we use the terminology ``NLO-NLL'' to indicate when,
at next-to-leading order, we include the LL and NLL soft-gluon terms 
(as well as scale dependent terms in $\delta(s_2)$). 
Also the terminology ``soft and virtual'' is used to denote
all the terms in Eq. (\ref{qgnlo}).
As we will see, we need to know both soft and virtual corrections at NLO
in order to derive the soft terms at NNLO with at least NNLL accuracy.

The NLO coefficients in Eq. (\ref{qgnlo}) are 
$c_3^{qg}=C_F+2C_A$,
\beq
c_2^{qg}=-\left(C_F + C_A\right) \ln\left(\frac{\mu_F^2}{Q_T^2}\right)
- {3\over 4} C_F - C_A \ln{\left(t u \over s Q_T^2\right)} \, ,
\eeq
and
\beq
c_1^{qg}=\frac{1}{2A^{qg}}\left[B_1^{qg}+B_2^{qg} n_f
+C_1^{qg}+C_2^{qg} n_f \right]+\frac{c_3^{qg}}{2}
\ln^2\left(\frac{Q_T^2}{Q^2}\right)
+c_2^{qg} \ln\left(\frac{Q_T^2}{Q^2}\right)\, , 
\eeq
with $C_A=N_c$, $n_f=5$ the number of light quark flavors, and
$B_1^{qg}$, $B_2^{qg}$, $C_1^{qg}$, and $C_2^{qg}$ 
as given in the Appendix
of Ref. \cite{gpw} but without the renormalization counterterms
and using $f_A \equiv\ln(A/Q^2)=0$. 

Note that we can write $c_2^{qg} \equiv c_{2 \; \mu}^{qg}+T_2^{qg}$
with $c_{2 \; \mu}^{qg} \equiv -(C_F+C_A)\ln(\mu_F^2/s)$.
Similarly we also write $c_1^{qg}\equiv c_{1 \; \mu}^{qg}+T_1^{qg}$ with
\beq
c_{1\; \mu}^{qg} \equiv \ln\left(\frac{\mu_F^2}{s}\right)
\left[-\frac{\beta_0}{4}+C_F\left(\ln\left(\frac{-u}{Q_T^2}\right)
-\frac{3}{4}\right)+C_A\ln\left(\frac{-t}{Q_T^2}\right)\right] 
+\frac{\beta_0}{4} \ln\left(\frac{\mu_R^2}{s}\right) \, ,
\eeq
where $\beta_0=(11C_A-2n_f)/3$ is the lowest-order beta function.
Thus,
$c_{2 \; \mu}^{qg}$ and $c_{1 \; \mu}^{qg}$ are scale-dependent parts of the
$c_2^{qg}$ and $c_1^{qg}$ coefficients, respectively, while
$T_2^{qg}$ and $T_1^{qg}$ are scale-independent parts.  
We have kept the factorization scale $\mu_F$ and the renormalization scale
$\mu_R$ separate.
Finally, another useful notation is
$c_1^{qg}\equiv {c'}_{1 \; \mu}^{qg}+{T'}_1^{qg}$
where ${c'}_{1 \; \mu}^{qg}$ is defined as $c_{1\; \mu}^{qg}$
with $Q_T^2$ instead of $s$ in the denominators of the logarithms
involving the scales $\mu_F$ and $\mu_R$.

Using the above conventions, the NNLO soft and virtual corrections for
$qg \longrightarrow Wq$ can be written as
\beq
E_Q\frac{d{\hat\sigma}^{(2)}_{qg \rightarrow Wq}}{d^3Q} = 
F^B_{qg \rightarrow Wq} 
\frac{\alpha_s^2(\mu_R^2)}{\pi^2} \, 
{\hat{\sigma'}}^{(2)}_{qg \rightarrow Wq}
\label{NNLOmqg}
\eeq
with
\beqa
{\hat{\sigma'}}^{(2)}_{qg \rightarrow Wq}&=& 
\frac{1}{2} (c_3^{qg})^2 \, \left[\frac{\ln^3(s_2/Q_T^2)}{s_2}\right]_+ 
+\left[\frac{3}{2} c_3^{qg} \, c_2^{qg} 
- \frac{\beta_0}{4} c_3^{qg}
+C_F \frac{\beta_0}{8}\right] \left[\frac{\ln^2(s_2/Q_T^2)}{s_2}\right]_+
\nonumber \\ && \hspace{-15mm}
{}+\left\{c_3^{qg} \, c_1^{qg} +(c_2^{qg})^2
-\zeta_2 \, (c_3^{qg})^2 -\frac{\beta_0}{2} \, T_2^{qg} 
+\frac{\beta_0}{4} c_3^{qg}  \ln\left(\frac{\mu_R^2}{s}\right)
+(C_F+C_A) \, K \right.
\nonumber \\ && \hspace{-15mm} \quad \quad \left.
{}+C_F \left[-\frac{K}{2} 
+\frac{\beta_0}{4} \, \ln\left(\frac{Q_T^2}{s}\right)\right]
-\frac{3}{16} \beta_0 C_F \right\}
\left[\frac{\ln(s_2/Q_T^2)}{s_2}\right]_+
\nonumber \\ && \hspace{-15mm} 
{}+\left\{c_2^{qg} \, c_1^{qg} -\zeta_2 \, c_2^{qg} \, c_3^{qg}
+\zeta_3 \, (c_3^{qg})^2 
-\frac{\beta_0}{2} T_1^{qg}
+\frac{\beta_0}{4}\, c_2^{qg} \ln\left(\frac{\mu_R^2}{s}\right) 
+{\cal G}_{qg}^{(2)}
\right. 
\nonumber \\ && \hspace{-15mm} \quad \quad
{}+(C_F+C_A) \left[\frac{\beta_0}{8} 
\ln^2\left(\frac{\mu_F^2}{s}\right)
-\frac{K}{2}\ln\left(\frac{\mu_F^2}{s}\right)\right]
-C_F\, K \, \ln\left(\frac{-u}{Q_T^2}\right)
-C_A\, K \, \ln\left(\frac{-t}{Q_T^2}\right)
\nonumber \\ && \hspace{-15mm} \quad \quad \left.
{}+C_F \, \left[\frac{\beta_0}{8}
\ln^2\left(\frac{Q_T^2}{s}\right)
-\frac{K}{2}\ln\left(\frac{Q_T^2}{s}\right)\right]
-\frac{3}{16}\beta_0 C_F\, \ln\left(\frac{Q_T^2}{s}\right) \right\}
\left[\frac{1}{s_2}\right]_+  
\nonumber \\ &&  \hspace{-15mm}
{}+\left\{\frac{1}{2}({c'}_{1 \; \mu}^{qg})^2
+{c'}_{1 \; \mu}^{qg} {T'}_{1}^{qg} 
+\frac{\beta_0}{4} {c'}_{1 \; \mu}^{qg} \ln\left(\frac{Q_T^2}{s}\right)
+\frac{\beta_0}{4} c_1^{qg} \ln\left(\frac{\mu_R^2}{Q_T^2}\right) 
-(C_F+C_A)^2 \frac{\zeta_2}{2} \ln^2\left(\frac{\mu_F^2}{Q_T^2}\right)
\right.
\nonumber \\ && \hspace{-15mm} \quad \quad
{}+(C_F+C_A) \ln\left(\frac{\mu_F^2}{Q_T^2}\right)
\left(\zeta_2 T_2^{qg}-\zeta_2 (C_F+C_A)\ln\left(\frac{Q_T^2}{s}\right)
-\zeta_3 c_3^{qg}\right)
\nonumber \\ && \hspace{-15mm} \quad \quad
{}-\frac{\beta_0^2}{32} \ln^2\left(\frac{\mu_R^2}{Q_T^2}\right)
-\frac{\beta_0^2}{16} \ln\left(\frac{\mu_R^2}{Q_T^2}\right)
\ln\left(\frac{Q_T^2}{s}\right)
+\frac{\beta_1}{16}\ln\left(\frac{\mu_R^2}{Q_T^2}\right)
\nonumber \\ && \hspace{-15mm} \quad \quad 
{}+\frac{\beta_0}{8} \left[\frac{3}{4}C_F+\frac{\beta_0}{4}
-C_F \, \ln\left(\frac{-u}{Q_T^2}\right)
-C_A \, \ln\left(\frac{-t}{Q_T^2}\right)\right] 
\left[\ln^2\left(\frac{\mu_F^2}{Q_T^2}\right)
+2\ln\left(\frac{\mu_F^2}{Q_T^2}\right)\ln\left(\frac{Q_T^2}{s}\right)\right]
\nonumber \\ && \hspace{-15mm} \quad \quad 
{}+C_F \frac{K}{2} \, \ln\left(\frac{-u}{Q_T^2}\right)
\ln\left(\frac{\mu_F^2}{Q_T^2}\right)
+C_A \frac{K}{2} \, \ln\left(\frac{-t}{Q_T^2}\right)
\ln\left(\frac{\mu_F^2}{Q_T^2}\right)
{}-({\gamma'}_{q/q}^{(2)}+{\gamma'}_{g/g}^{(2)}) 
\ln\left(\frac{\mu_F^2}{Q_T^2}\right)
\nonumber \\ && \hspace{-15mm} \left. \quad \quad 
{}+ R_{qg}^{(2)} \right\} \delta(s_2) \, ,
\label{NNLOqg}
\eeqa
where $K=C_A(67/18-\zeta_2)-5n_f/9$ is a two-loop function
in the $\overline{\rm MS}$ scheme \cite{KodTr}, $\zeta_2=\pi^2/6$ and
$\zeta_3=1.2020569\cdots$ are Riemann zeta functions, 
$\beta_1=34 C_A^2/3 \, - \, 2n_f(C_F+5C_A/3)$ is the 
next-to-leading order beta function, and
\beq
{\gamma'}_{q/q}^{(2)}=C_F^2\left(\frac{3}{32}-\frac{3}{4}\zeta_2
+\frac{3}{2}\zeta_3\right)
+C_F C_A\left(-\frac{3}{4}\zeta_3+\frac{11}{12}\zeta_2+\frac{17}{96}\right)
+n_f C_F \left(-\frac{\zeta_2}{6}-\frac{1}{48}\right)\, ,
\eeq
\beq
{\gamma'}_{g/g}^{(2)}=C_A^2\left(\frac{2}{3}+\frac{3}{4}\zeta_3\right)
-n_f\left(\frac{C_F}{8}+\frac{C_A}{6}\right) \, , 
\eeq
are two-loop parton anomalous dimensions \cite{GALY,GFP}.

The soft corrections are the $[\ln^3(s_2/Q_T^2)/s_2]_+$ term (which is the 
LL, since $n=2$, $m=3$ in Eq.~(\ref{zplus})),
$[\ln^2(s_2/Q_T^2)/s_2]_+$ term (NLL, $n=2$, $m=2$),
$[\ln(s_2/Q_T^2)/s_2]_+$ term (NNLL, $n=2$, $m=1$), and
$[1/s_2]_+$ term (NNNLL, $n=2$, $m=0$).
The function ${\cal G}^{(2)}_{q g}$ in the NNNLL term denotes a set of 
two-loop contributions \cite{NKJO2,NKuni} and is given by
\beqa
{\cal G}^{(2)}_{q g}&=&C_F^2\left(-\frac{3}{32}+\frac{3}{4}\zeta_2
-\frac{3}{2}\zeta_3\right)+ C_F C_A \left(\frac{3}{4} \zeta_3
-\frac{11}{12}\zeta_2-\frac{189}{32}\right)
\nonumber \\ && \hspace{-5mm}
{}+C_A^2 \left(\frac{7}{4} \zeta_3
+\frac{11}{3}\zeta_2-\frac{41}{216}\right)
+ n_f C_F \left(\frac{1}{6}\zeta_2+\frac{17}{16}\right) 
+n_f C_A \left(-\frac{2}{3}\zeta_2
-\frac{5}{108}\right) \, .
\eeqa
Note that we have not included in ${\cal G}^{(2)}_{q g}$ two-loop
process-dependent contributions; however from related studies for
other processes, including top hadroproduction \cite{NKRV} 
and direct photon production \cite{NKJO2}
we expect such contributions to be small. It is actually the 
$-\zeta_2 c_2^{qg} c_3^{qg}+\zeta_3 (c_3^{qg})^2$ terms
in Eq.~(\ref{NNLOqg}) that provide the major contribution to the NNNLL term.

We use the terminology ``NNLO-NNLL'' below, to indicate that we include
the LL, NLL, and NNLL terms at NNLO, while we use ``NNLO-NNNLL'' to
indicate that in addition we include the NNNLL terms as well. 

The term proportional to $\delta(s_2)$ includes the virtual corrections.
We remind the reader that here we do not calculate the full virtual 
corrections; the term $R_{qg}^{(2)}$ that denotes the scale-independent
virtual corrections is currently unknown.
However, we have calculated explicitly here all the 
$\delta(s_2)$ terms that  include the 
scale-dependence. Note that when discussing scale dependence 
at NNLO-NNNLL we include all scale-dependent $\delta(s_2)$ terms,
which is a consistent approach from the
resummation procedure, see Ref. \cite{NKtop}. 

\subsection{The $q{\bar q} \longrightarrow Wg$ subprocess}

Next, we consider the $q{\bar q} \longrightarrow Wg$
partonic subprocess.
Here the Born differential cross section is
\beq
E_Q \frac{d\sigma^B_{q {\bar q}\rightarrow Wg}}{d^3Q}
=F^B_{q{\bar q} \rightarrow Wg} \, \delta(s_2) \, ,
\eeq
where
\beqa
F^B_{q{\bar q} \rightarrow Wg} &=&\frac{\alpha \alpha_s(\mu_R^2)C_F}{sN_c}
A^{q\bar q}\,  |L_{f_bf_a}|^2 \, , \\
A^{q\bar q} &=& \frac{u}{t}+\frac{t}{u}+\frac{2Q^2s}{tu} \, .
\nonumber
\eeqa 

The NLO soft and virtual corrections in single-particle inclusive kinematics
can be written as 
\beqa
E_Q\frac{d{\hat\sigma}^{(1)}_{q{\bar q} \rightarrow Wg}}{d^3Q} &=&
{F^B_{q{\bar q} \rightarrow Wg}} {\alpha_s(\mu_R^2)\over\pi}\,
\left\{c_3^{q \bar q} \, \left[\frac{\ln(s_2/Q_T^2)}{s_2}\right]_+ 
+c_2^{q \bar q} \, \left[\frac{1}{s_2}\right]_+ 
+c_1^{q \bar q} \, \delta(s_2)\right\} \, .
\label{qqbarnlo}
\eeqa
Here the NLO coefficients are
$c_3^{q \bar q}=4C_F-C_A$,
\beq
c_2^{q \bar q}=- 2 C_F \ln\left(\frac{\mu_F^2}{Q_T^2}\right) 
- \left(2 C_F- C_A \right) \ln\left(\frac{t u}{s Q_T^2}\right) 
-\frac{\beta_0}{4}
\eeq
and
\beq
c_1^{q \bar q}=\frac{1}{2A^{q \bar q}}\left[B_1^{q \bar q}+C_1^{q \bar q}
+(B_2^{q \bar q}+D_{aa}^{(0)}) \, n_f \right] 
+\frac{c_3^{q \bar q}}{2}
\ln^2\left(\frac{Q_T^2}{Q^2}\right)
+c_2^{q \bar q} \ln\left(\frac{Q_T^2}{Q^2}\right)\, ,
\eeq
with $B_1^{q \bar q}$, $B_2^{q \bar q}$,
$C_1^{q \bar q}$, and $D_{aa}^{(0)}$
as given in the Appendix of Ref. \cite{gpw} but without the renormalization 
counterterms and using $f_A=0$.
Again, we can write $c_1^{q \bar q}=c_{1 \; \mu}^{q \bar q}+T_1^{q \bar q}$
with
\beq
c_{1 \; \mu}^{q \bar q}=\ln\left(\frac{\mu_F^2}{s}\right)  
C_F\left[\ln\left(\frac{tu}{Q_T^4}\right)-\frac{3}{2}\right]
+\frac{\beta_0}{4} \ln\left(\frac{\mu_R^2}{s}\right) \, ,
\eeq
and $c_2^{q \bar q}=c_{2 \; \mu}^{q \bar q}+T_2^{q \bar q}$ 
with $c_{2 \; \mu}^{q \bar q}=- 2 C_F \ln(\mu_F^2/s)$. 
Finally, another useful notation is
$c_1^{q \bar q}\equiv {c'}_{1 \; \mu}^{q \bar q}+{T'}_1^{q \bar q}$
where ${c'}_{1 \; \mu}^{q \bar q}$ is defined as $c_{1\; \mu}^{q \bar q}$
with $Q_T^2$ instead of $s$ in the denominators of the logarithms
involving the scales $\mu_F$ and $\mu_R$.

The NNLO soft and virtual corrections 
for $q{\bar q} \longrightarrow Wg$ can be written as
\beq
E_Q\frac{d{\hat\sigma}^{(2)}_{q{\bar q} \rightarrow Wg}}{d^3Q} =
{F^B_{q{\bar q} \rightarrow Wg}}
\frac{\alpha_s^2(\mu_R^2)}{\pi^2} \, {\hat{\sigma'}}^{(2)}_{q{\bar q} 
\rightarrow Wg}
\label{NNLOmqqb}
\eeq
with
\beqa
{\hat{\sigma'}}^{(2)}_{q{\bar q} \rightarrow Wg}&=& 
\frac{1}{2} (c_3^{q \bar q})^2 \, \left[\frac{\ln^3(s_2/Q_T^2)}{s_2}\right]_+ 
+\left[\frac{3}{2} c_3^{q \bar q} \, c_2^{q \bar q} 
- \frac{\beta_0}{4} c_3^{q \bar q}
+C_A \frac{\beta_0}{8}\right] \left[\frac{\ln^2(s_2/Q_T^2)}{s_2}\right]_+ 
\nonumber \\ && \hspace{-10mm}
{}+\left\{c_3^{q \bar q} \, c_1^{q \bar q} +(c_2^{q \bar q})^2
-\zeta_2 \, (c_3^{q \bar q})^2 -\frac{\beta_0}{2} \, T_2^{q \bar q} 
+\frac{\beta_0}{4} c_3^{q \bar q}  \ln\left(\frac{\mu_R^2}{s}\right)
+2C_F \, K \right.
\nonumber \\ && \hspace{-10mm} \quad \quad \left.
{}+C_A \left[-\frac{K}{2} 
+\frac{\beta_0}{4} \, \ln\left(\frac{Q_T^2}{s}\right)\right]
-\frac{\beta_0^2}{16} \right\}
\left[\frac{\ln(s_2/Q_T^2)}{s_2}\right]_+ 
\nonumber \\ && \hspace{-10mm} 
{}+\left\{c_2^{q \bar q} \, c_1^{q \bar q} -\zeta_2 \, c_2^{q \bar q}
 \, c_3^{q \bar q}+\zeta_3 \, (c_3^{q \bar q})^2 
-\frac{\beta_0}{2} T_1^{q \bar q}
+\frac{\beta_0}{4}\, c_2^{q \bar q} \ln\left(\frac{\mu_R^2}{s}\right) 
+{\cal G}^{(2)}_{q \bar q} \right. 
\nonumber \\ && \hspace{-10mm} \quad \quad
{}+C_F \left[\frac{\beta_0}{4} 
\ln^2\left(\frac{\mu_F^2}{s}\right)
-K\ln\left(\frac{\mu_F^2}{s}\right)
-K \, \ln\left(\frac{tu}{Q_T^4}\right)\right]
\nonumber \\ && \hspace{-10mm} \quad \quad \left.
{}+C_A \, \left[\frac{\beta_0}{8}
\ln^2\left(\frac{Q_T^2}{s}\right)
-\frac{K}{2}\ln\left(\frac{Q_T^2}{s}\right)\right]
-\frac{\beta_0^2}{16}\, \ln\left(\frac{Q_T^2}{s}\right) \right\}
\left[\frac{1}{s_2}\right]_+  
\nonumber \\ &&  \hspace{-10mm}
{}+\left\{\frac{1}{2}({c'}_{1 \; \mu}^{q \bar q})^2
+{c'}_{1 \; \mu}^{q \bar q} {T'}_{1}^{q \bar q}
+\frac{\beta_0}{4}{c'}_{1 \; \mu}^{q \bar q}
\ln\left(\frac{Q_T^2}{s}\right)  
+\frac{\beta_0}{4} c_1^{q \bar q} \ln\left(\frac{\mu_R^2}{Q_T^2}\right) 
\right.
\nonumber \\ && \hspace{-10mm} \quad \quad
{}-2C_F^2 \zeta_2 \ln^2\left(\frac{\mu_F^2}{Q_T^2}\right)
+2C_F \ln\left(\frac{\mu_F^2}{Q_T^2}\right)
\left(\zeta_2 T_2^{q \bar q}-2 \zeta_2 C_F\ln\left(\frac{Q_T^2}{s}\right)
-\zeta_3 c_3^{q \bar q}\right)
\nonumber \\ && \hspace{-10mm} \quad \quad
{}-\frac{\beta_0^2}{32} \ln^2\left(\frac{\mu_R^2}{Q_T^2}\right)
-\frac{\beta_0^2}{16} \ln\left(\frac{\mu_R^2}{Q_T^2}\right)
\ln\left(\frac{Q_T^2}{s}\right)
+\frac{\beta_1}{16}\ln\left(\frac{\mu_R^2}{Q_T^2}\right)
\nonumber \\ && \hspace{-10mm} \quad \quad 
{}+\frac{\beta_0}{8} \left[\frac{3}{2}C_F
-C_F \, \ln\left(\frac{tu}{Q_T^4}\right)\right] 
\left[\ln^2\left(\frac{\mu_F^2}{Q_T^2}\right)
+2\ln\left(\frac{\mu_F^2}{Q_T^2}\right)\ln\left(\frac{Q_T^2}{s}\right)\right]
\nonumber \\ && \hspace{-10mm} \left. \quad \quad 
{}+C_F \frac{K}{2} \, \ln\left(\frac{tu}{Q_T^4}\right)
\ln\left(\frac{\mu_F^2}{Q_T^2}\right)
-2{\gamma'}_{q/q}^{(2)} \ln\left(\frac{\mu_F^2}{Q_T^2}\right)
+ R_{q{\bar q}}^{(2)} \right\} \delta(s_2) \, ,
\label{NNLOqqb}
\eeqa
The function ${\cal G}^{(2)}_{q {\bar q}}$ denotes again
a set of two-loop contributions \cite{NKJO2,NKuni} and is given by
\beq
{\cal G}^{(2)}_{q {\bar q}}
=C_F C_A \left(\frac{7}{2} \zeta_3
+\frac{22}{3}\zeta_2-\frac{299}{27}\right)
+n_f C_F \left(-\frac{4}{3}\zeta_2+\frac{50}{27}\right) \, . 
\eeq
Again, we have not included in ${\cal G}^{(2)}_{q {\bar q}}$
two-loop process-dependent contributions.

We also note that we do not calculate the full virtual corrections.
All $\delta(s_2)$ terms shown here only include the scale-dependence
and are used in the study of the scale dependence of the NNLO-NNNLL
cross section. The term $R_{q{\bar q}}^{(2)}$ is unknown.

\mysection{Numerical results}

For the $W$ hadroproduction cross section for
$ h_A (P_A) + h_B (P_B) \rightarrow W (Q) + X$
we defined the hadronic and partonic kinematical invariants in Section 2.
The hadronic kinematical variables $T$ and $U$ can be written 
alternatively as
$T = m_W^2 - m_T \sqrt{S} e^{-y}$ and
$U = m_W^2 - m_T \sqrt{S} e^{y}$.
Here $m_W$ is the $W$ mass, $m_T = \sqrt{Q_T^2 + m_W^2}$ is 
the transverse mass,
and $y$ is the rapidity of the vector boson. The differential $Q_T$
distribution can then be written as
\begin{eqnarray} \frac{d \sigma_{h_A h_B\rightarrow W+X}}{d Q_T^2}
\left(S,m_W^2,Q_T\right) &=& \sum_f \int_0^1 dy' \int^1_A dx_1
\int^{s_2^{\rm max}}_0 d s_2 \frac{2 \pi Y}{x_1 S-\sqrt{S} m_T e^y} \\
&\times& \phi_{f_a/h_A}\left(x_1,\mu_F^2\right)
\phi_{f_b/h_B}\left(x_2,\mu_F^2\right)
E_Q \frac{d{\hat \sigma}_{f_a f_b\rightarrow W+ X}}{d^3 Q}
\left(x_1,x_2,y\right) \nonumber,
\end{eqnarray}
where $Y = \ln{\left(B+\sqrt{B^2-1}\right)}$, 
$B = (S+m_W^2)/(2 m_T \sqrt{S})$,
$y= Y (2 y'-1)$,
\beqa
s_2^{\rm max} &=& m_W^2-\sqrt{S}m_T e^y + x_1
\left(S-\sqrt{S}m_T e^{-y}\right),\\
A &=& \frac{\sqrt{S}m_T e^y -m_W^2}{S-\sqrt{S} m_T e^{-y}},\quad
x_2 = \frac{s_2-m_W^2+\sqrt{S}m_T x_1 e^{-y}}{x_1 S - \sqrt{S} m_T e^y}.
\eeqa

\begin{figure}[htb] 
\setlength{\epsfxsize=0.85\textwidth}
\setlength{\epsfysize=0.55\textheight}
\centerline{\epsffile{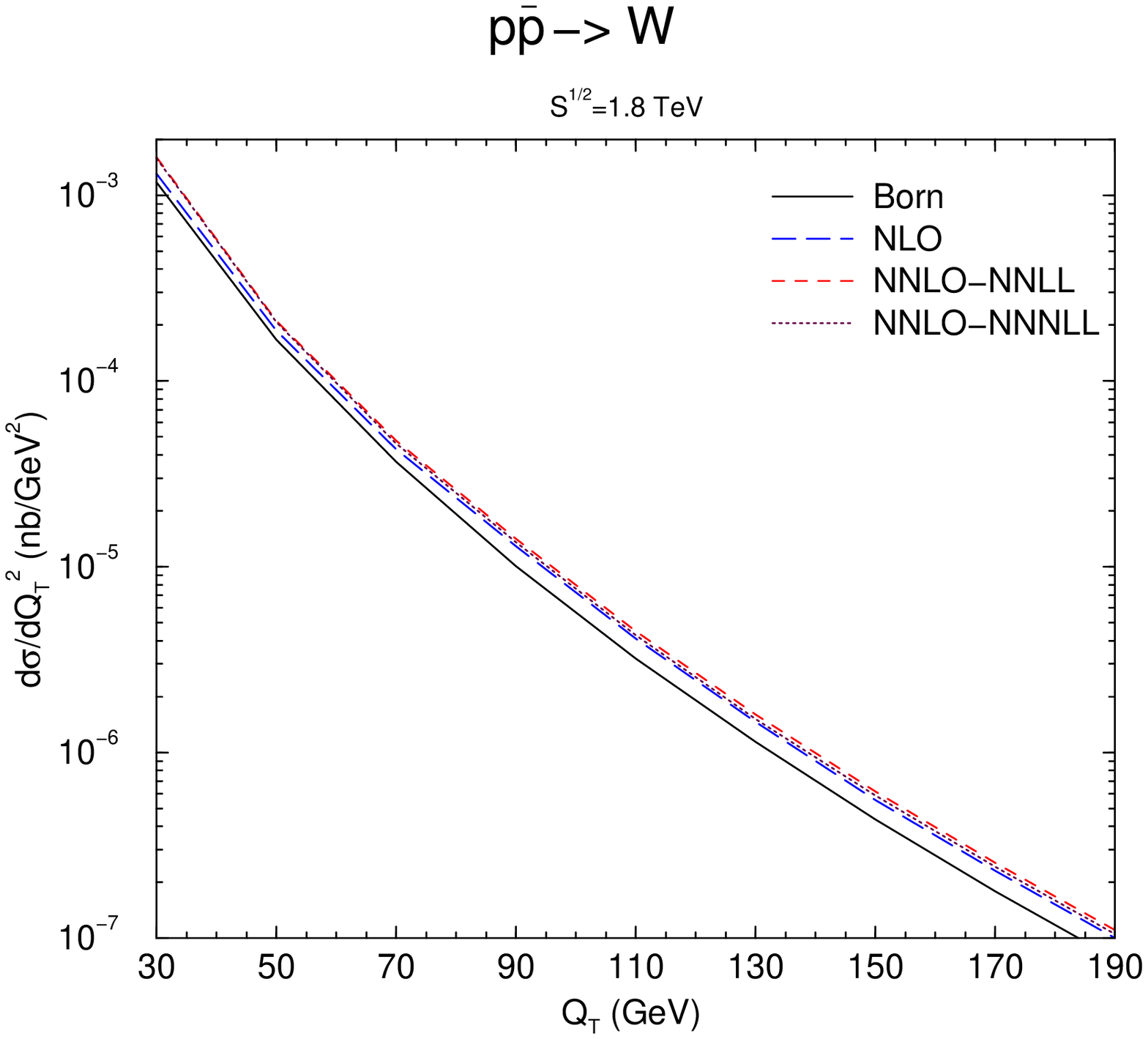}}
\caption[]{The differential cross section,
$d\sigma/dQ_T^2$, for $W$ hadroproduction in $p \bar p$ collisions
at the Tevatron with $\sqrt{S}=1.8$ TeV and $\mu_F=\mu_R=Q_T$.
Shown are the Born (solid line), NLO (long-dashed line),
NNLO-NNLL (short-dashed line), and NNLO-NNNLL (dotted line) results.
}
\label{fig1} 
\end{figure}

\begin{figure}[htb] 
\setlength{\epsfxsize=0.85\textwidth}
\setlength{\epsfysize=0.55\textheight}
\centerline{\epsffile{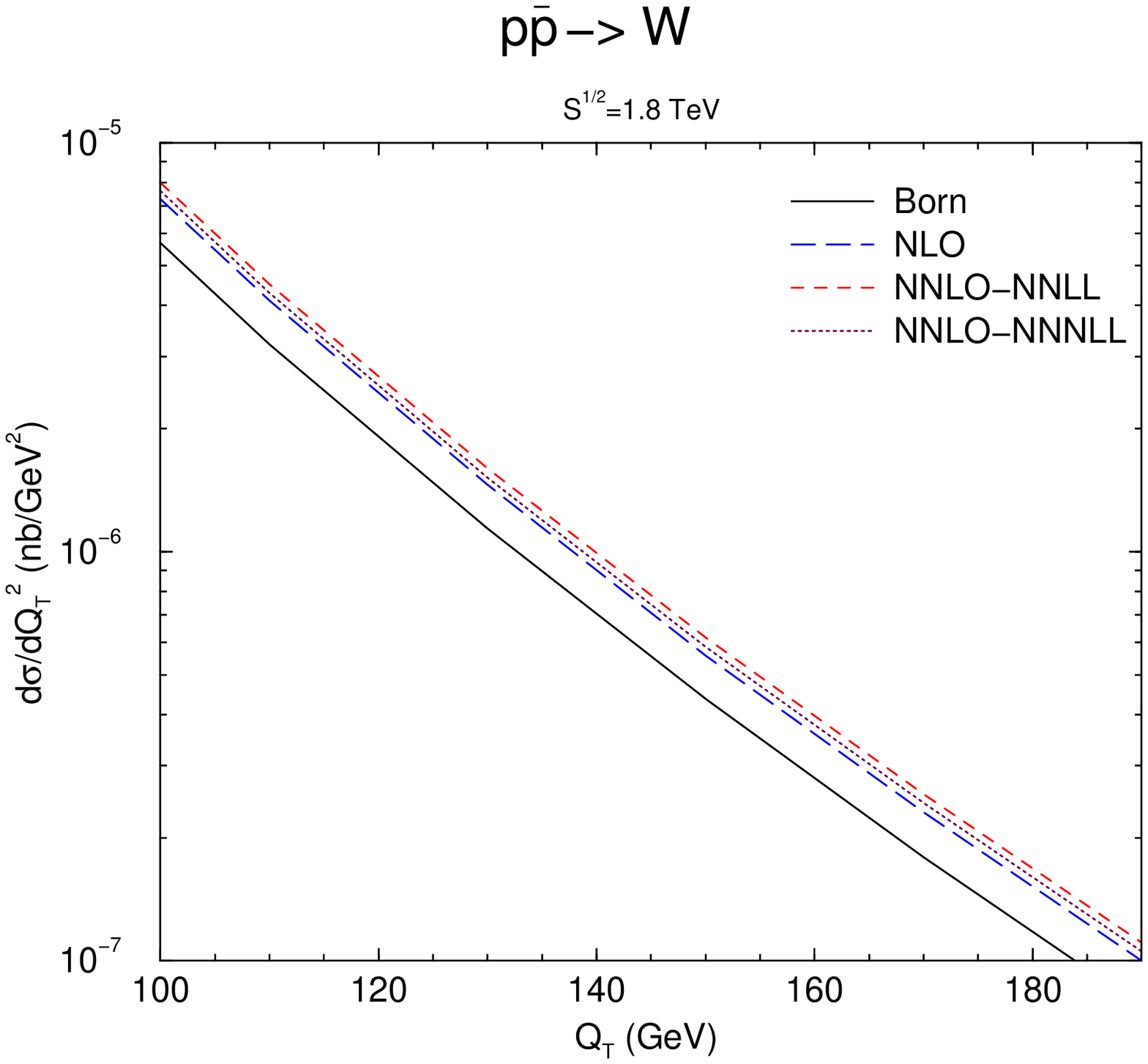}}
\caption[]{The differential cross section,
$d\sigma/dQ_T^2$, of Fig. 1 at high $Q_T$. The labels are
the same as in Fig. 1.}
\label{fig2} 
\end{figure}

We now apply our results to $W$ hadroproduction at large transverse momentum
at the Tevatron. Throughout we use the MRST2002 approximate NNLO
parton densities \cite{MRST}.
In Fig. 1 we plot the transverse momentum distribution,
$d\sigma/dQ_T^2$, for $W$ hadroproduction at the Tevatron Run I 
with $\sqrt{S}=1.8$ TeV. Here we have set $\mu_F=\mu_R=Q_T$.
We plot Born, exact NLO \cite{AR,gpw}, NNLO-NNLL, and 
NNLO-NNNLL results. We see that the 
NLO corrections provide a significant enhancement of the Born
cross section. The NNLO-NNLL corrections provide a further
modest enhancement of the $Q_T$ distribution. 
If we increase the accuracy by including the NNNLL contributions,
which are negative,
then we find that the NNLO-NNNLL cross section lies between
the NLO and NNLO-NNLL results. Since it is hard to distinguish
between the curves in Fig. 1, we provide another figure, Fig. 2, 
which emphasizes
the high-$Q_T$ region where the soft-gluon approximation holds best.

\begin{figure}[htb] 
\setlength{\epsfxsize=0.85\textwidth}
\setlength{\epsfysize=0.55\textheight}
\centerline{\epsffile{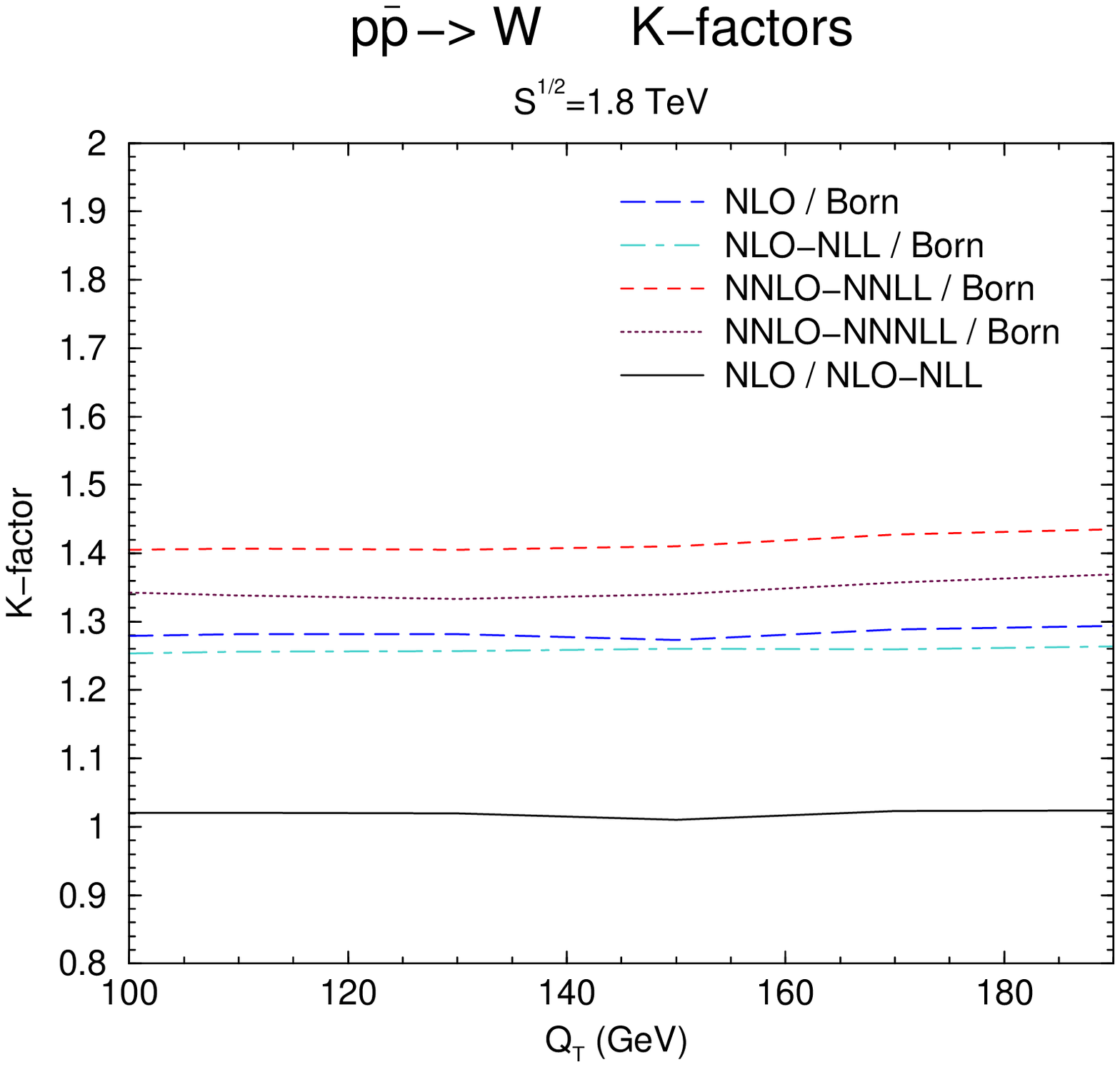}}
\caption[]{The $K$-factors for the differential cross section,
$d\sigma/dQ_T^2$,  for $W$ hadroproduction 
in $p \bar p$ collisions at the Tevatron with $\sqrt{S}=1.8$ TeV 
and $\mu_F=\mu_R=Q_T$.
Shown are the $K$- factors for exact NLO/Born (long-dashed line), 
NLO-NLL/Born (dash-dotted line), NNLO-NNLL/Born (short-dashed line),
and approximate NNLO-NNNLL/Born (dotted line) results.
Also shown is the ratio of the exact NLO to the NLO-NLL cross section
(solid line).
}
\label{fig3} 
\end{figure}

In Fig. 3 we plot the $K$-factors, i.e. the ratios of
cross sections at various orders and accuracies
to the Born cross section, all with $\mu_F=\mu_R=Q_T$,
in the high-$Q_T$ region.
We also show the ratio of the exact NLO to the NLO-NLL cross section.
It is clear from this line being very close to 1 that the NLO-NLL result
is a very good approximation to the full NLO result, i.e. the
soft-gluon corrections overwhelmingly dominate the NLO cross section. 
The difference between NLO and NLO-NLL is only 2\% for $Q_T > 90$ GeV
and less than 10\% for lower $Q_T$ down to 30 GeV. The fact that the
soft-gluon corrections dominate the NLO cross section
is a major justification for studying the NNLO soft gluon corrections
to this process. We can also see that the various $K$-factors shown in Fig. 3 
are moderate, and nearly constant over the $Q_T$ range shown even though
the distributions themselves span two orders of magnitude in this range.
The NLO corrections are nearly 30\% over the Born, with the NNLO
corrections giving an additional increase, so that the NNLO-NNLL/Born
$K$-factor is around 1.4 and the NNLO-NNNLL/Born $K$-factor is around
1.35.  

\begin{figure}[htb] 
\setlength{\epsfxsize=0.85\textwidth}
\setlength{\epsfysize=0.55\textheight}
\centerline{\epsffile{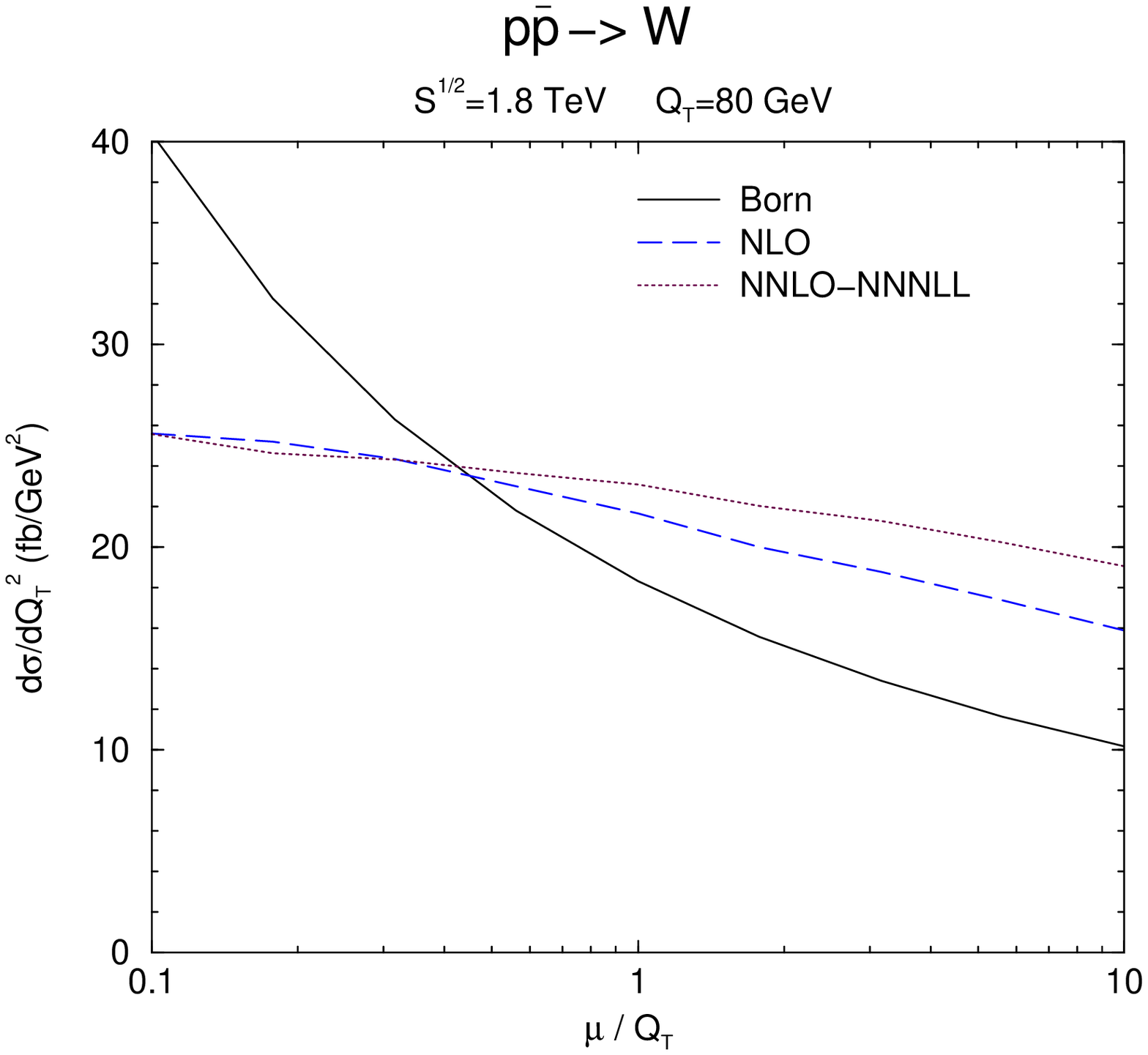}}
\caption[]{The differential cross section,
$d\sigma/dQ_T^2$, for $W$ hadroproduction in $p \bar p$ collisions
at the Tevatron with $\sqrt{S}=1.8$ TeV, $Q_T=80$ GeV, and 
$\mu \equiv \mu_F=\mu_R$.
Shown are the Born (solid line), exact NLO (long-dashed line), 
and NNLO-NNNLL (dotted line) results.
}
\label{fig4} 
\end{figure}

In Fig. 4 we plot the scale dependence of the differential cross section
for $Q_T=80$ GeV. We define $\mu \equiv \mu_F=\mu_R$ and plot
the differential cross section versus $\mu/Q_T$ over two
orders of magnitude: $0.1 < \mu/Q_T < 10$. We note the good
stabilization of the cross section when the NLO corrections are
included, and the further improvement when the NNLO-NNNLL corrections
(which include all the soft and virtual NNLO scale terms) are added.
The NNLO-NNNLL result approaches the scale independence expected of a truly
physical cross section. 

\begin{figure}[htb] 
\setlength{\epsfxsize=0.85\textwidth}
\setlength{\epsfysize=0.55\textheight}
\centerline{\epsffile{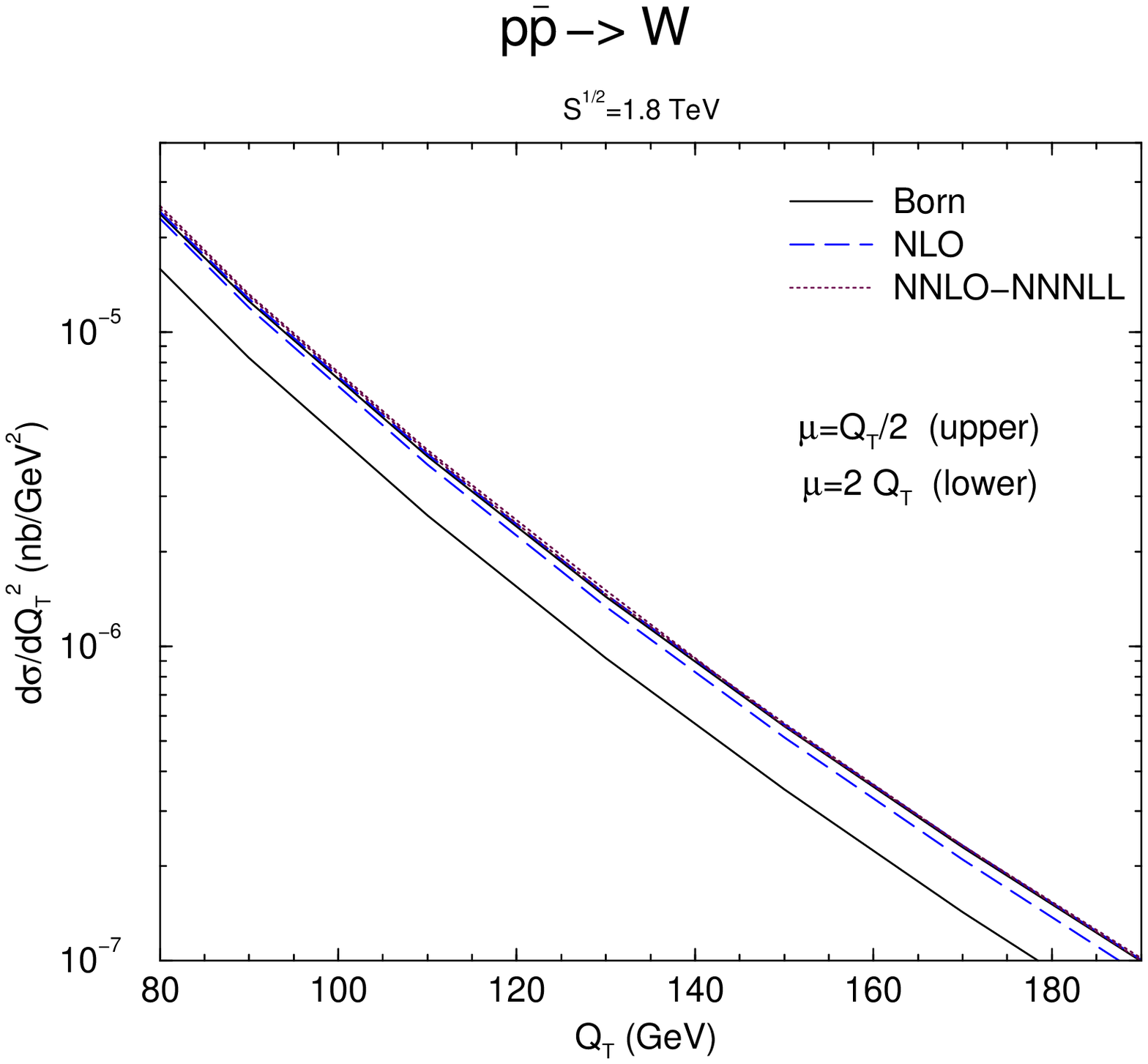}}
\caption[]{The differential cross section,
$d\sigma/dQ_T^2$, for $W$ hadroproduction in $p \bar p$ collisions
at the Tevatron with $\sqrt{S}=1.8$ TeV and 
$\mu\equiv\mu_F=\mu_R=Q_T/2$ or $2Q_T$.
Shown are the Born (solid lines), NLO (long-dashed lines), 
and NNLO-NNNLL (dotted lines) results. The upper lines
are with $\mu=Q_T/2$, the lower lines with $\mu=2 Q_T$.
}
\label{fig5} 
\end{figure}

In Fig. 5 we plot the differential cross section
$d\sigma/dQ_T^2$ at high $Q_T$ with $\sqrt{S}=1.8$ TeV
for two values of scale, $Q_T/2$ and $2Q_T$,
often used to display the uncertainty due to scale variation.
We note that while the variation of the Born cross section is
significant, the variation at NLO is much smaller, and at
NNLO-NNNLL it is very small. In fact the two NNLO-NNNLL curves
lie on top of the $\mu=Q_T/2$ Born curve, and so does the
NLO $\mu=Q_T/2$ curve. These results are consistent with 
Fig. 4.

\begin{figure}[htb] 
\setlength{\epsfxsize=0.85\textwidth}
\setlength{\epsfysize=0.55\textheight}
\centerline{\epsffile{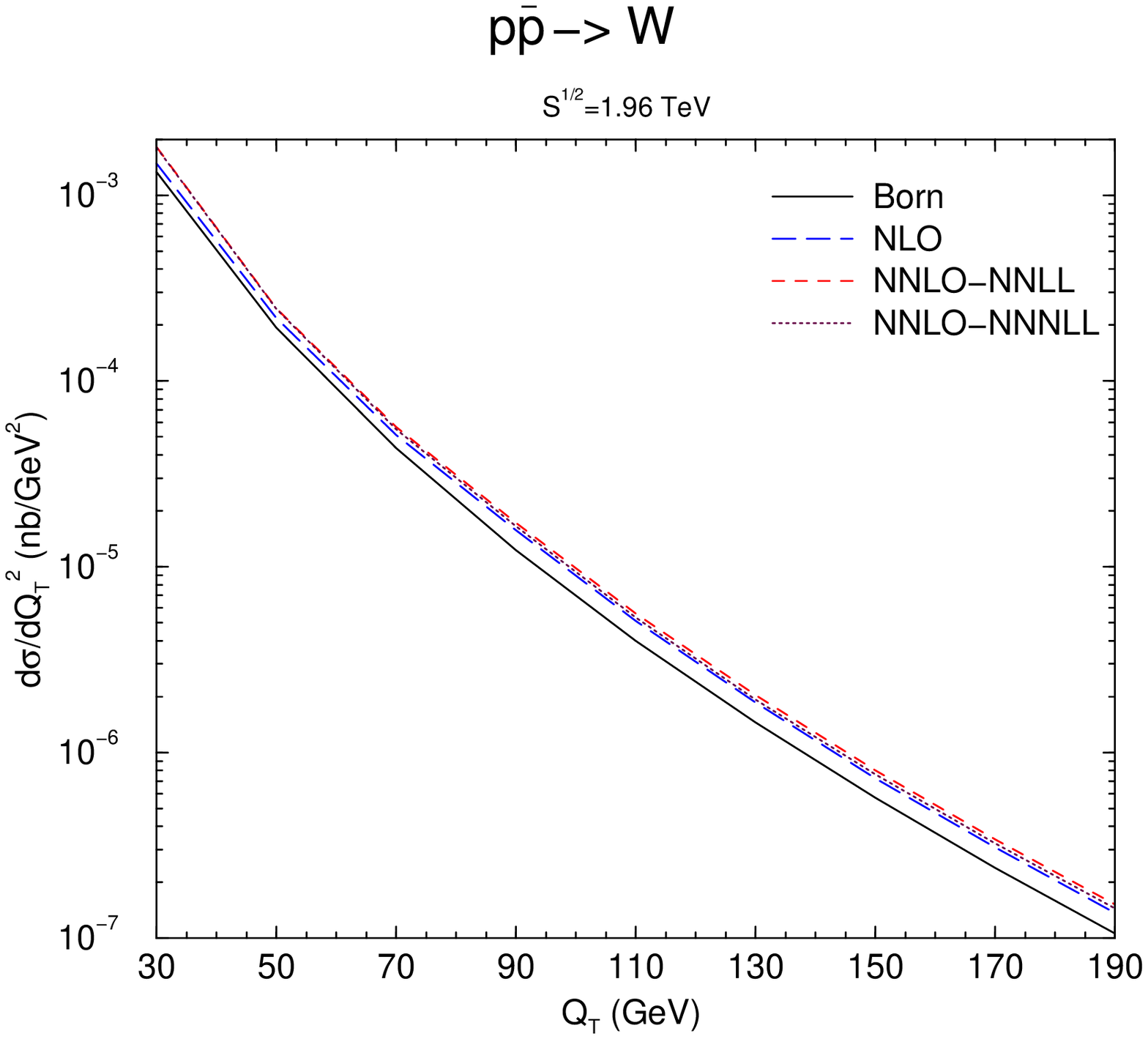}}
\caption[]{The differential cross section,
$d\sigma/dQ_T^2$, for $W$ hadroproduction in $p \bar p$ collisions
at the Tevatron Run II with $\sqrt{S}=1.96$ TeV and $\mu_F=\mu_R=Q_T$.
Shown are the Born (solid line), exact NLO (long-dashed line), 
NNLO-NNLL (short-dashed line), and NNLO-NNNLL (dotted line) results.
}
\label{fig6} 
\end{figure}

\begin{figure}[htb] 
\setlength{\epsfxsize=0.85\textwidth}
\setlength{\epsfysize=0.55\textheight}
\centerline{\epsffile{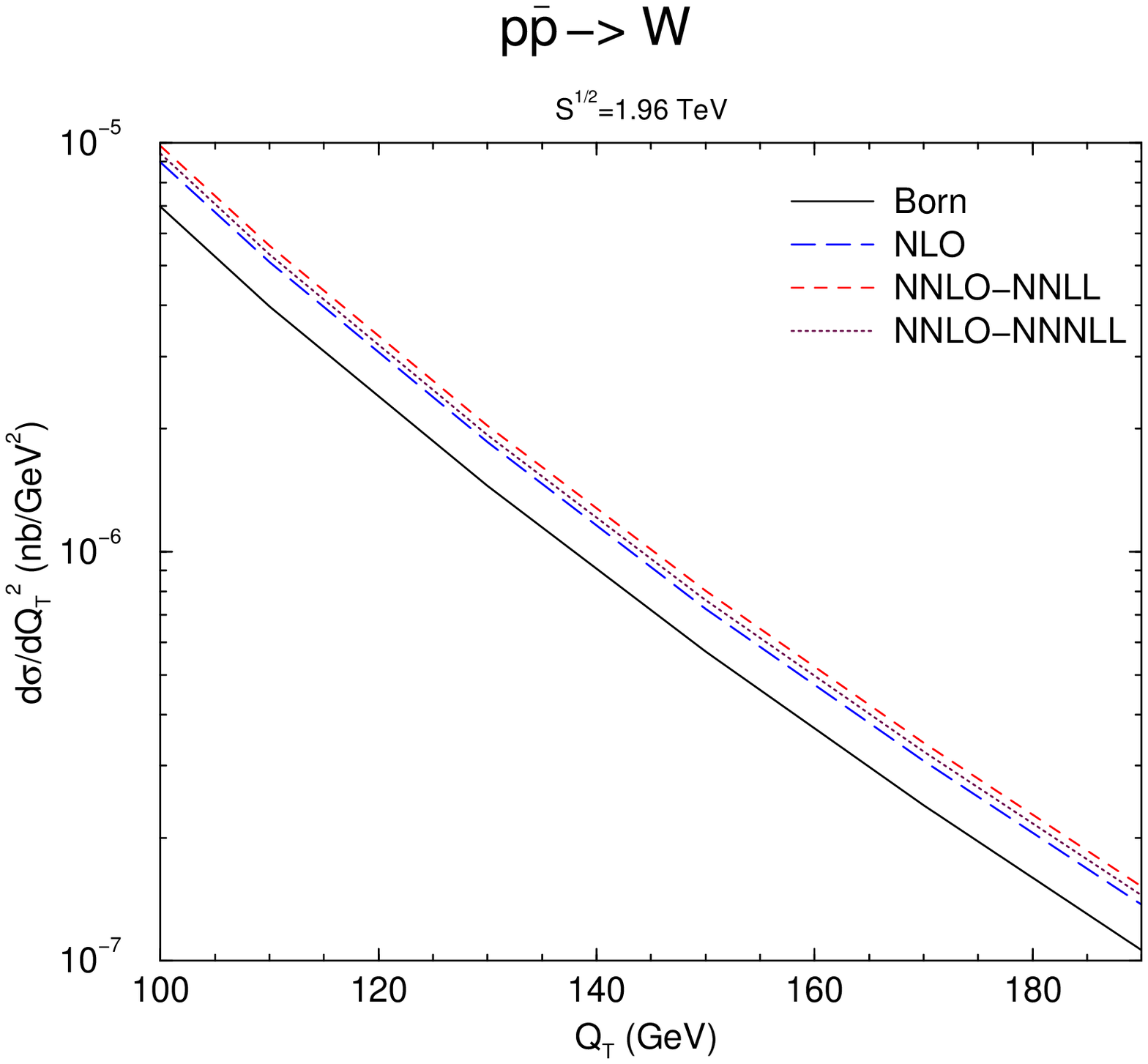}}
\caption[]{The differential cross section,
$d\sigma/dQ_T^2$, of Fig. 6 at high $Q_T$. The labels are
the same as in Fig. 6.
}
\label{fig7} 
\end{figure}

In Fig. 6 we plot the transverse momentum distribution,
$d\sigma/dQ_T^2$, for $W$ hadroproduction at the Tevatron 
Run II with $\sqrt{S}=1.96$ TeV. Again, we have set $\mu_F=\mu_R=Q_T$.
The relative size of the corrections is similar to Fig. 1. 
In Fig. 7 we display the high-$Q_T$ region more clearly. 
We do not plot the $K$-factors for Run II as they are nearly identical to
those for $\sqrt{S}=1.8$ TeV.

\begin{figure}[htb] 
\setlength{\epsfxsize=0.85\textwidth}
\setlength{\epsfysize=0.55\textheight}
\centerline{\epsffile{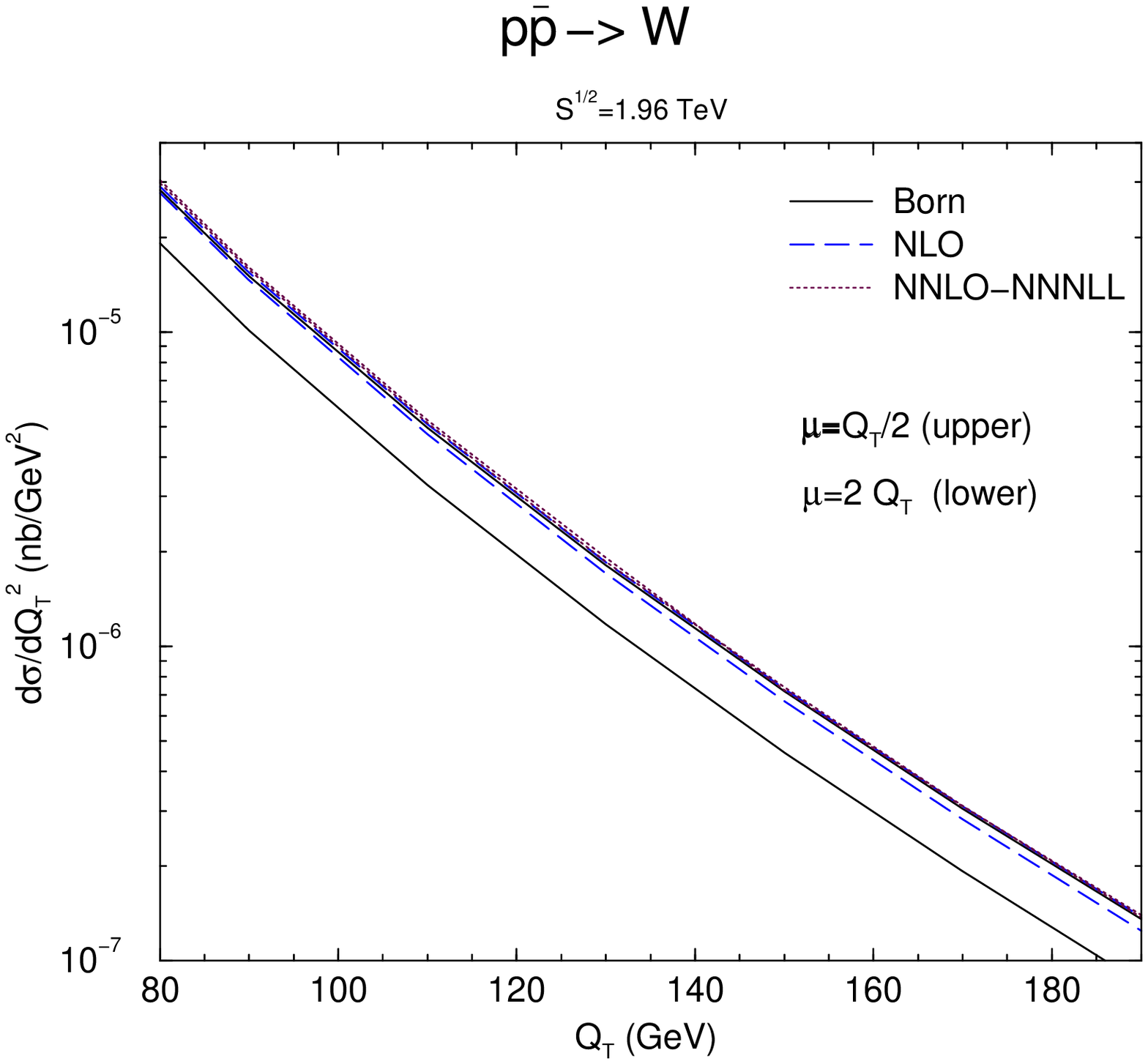}}
\caption[]{The differential cross section,
$d\sigma/dQ_T^2$, for $W$ hadroproduction in $p \bar p$ collisions
at the Tevatron with $\sqrt{S}=1.96$ TeV and 
$\mu \equiv \mu_F=\mu_R=Q_T/2$ or $2Q_T$.
Shown are the Born (solid lines), NLO (long-dashed lines), 
and NNLO-NNNLL (dotted lines) results. The upper lines
are with $\mu=Q_T/2$, the lower lines with $\mu=2 Q_T$.
}
\label{fig8} 
\end{figure}

In Fig. 8 we plot the differential cross section
$d\sigma/dQ_T^2$ at high $Q_T$ with $\sqrt{S}=1.96$ TeV
for two values of scale, $Q_T/2$ and $2Q_T$.
The results are analogous to those in Fig. 5.

\mysection{Conclusion}

We have presented the NNLO soft-gluon corrections 
for $W$ hadroproduction at large transverse momentum
in $p{\bar p}$ collisions, in particular at the Tevatron Run I and II.
We have shown that the NLO soft-gluon corrections completely
dominate the NLO differential cross section at large transverse momentum,
while the NNLO soft-gluon corrections provide modest enhancements
and further decrease the factorization and renormalization
scale dependence of the transverse momentum distributions.

\mysection*{Acknowledgements}

We are grateful to Richard Gonsalves for providing
us with numerical results for the exact NLO corrections.
The research of N.K. has been supported by a Marie Curie Fellowship of 
the European Community programme ``Improving Human Research Potential'' 
under contract number HPMF-CT-2001-01221.
A.S.V.~thanks the II.~Institut f{\" u}r Theoretische Physik at the
University of Hamburg for hospitality and acknowledges the support of
PPARC (Postdoctoral Fellowship: PPA/P/S/1999/00446).

\end{document}